\documentclass[prl,twocolumn,showpacs,preprintnumbers,superscriptaddress]{revtex4}

\usepackage{amsmath}
\usepackage{graphicx}

\usepackage{amsmath}
\usepackage{amssymb}
\usepackage{graphicx}
\usepackage[british]{babel}

\begin{document}

\preprint{\today.}

\title{Bulk-mediated surface diffusion on a cylinder: propagators and
crossovers}

\author{Aleksei V. Chechkin}
\affiliation{Institute for Theoretical Physics NSC KIPT,
Akademicheskaya st.1, 61108 Kharkov, Ukraine}
\affiliation{Physik Department, Technical University of Munich,
James Franck Strasse, 85747 Garching, Germany}
\author{Irwin M. Zaid}
\affiliation{Physik Department, Technical University of Munich,
James Franck Strasse, 85747 Garching, Germany}
\author{Michael A. Lomholt}
\affiliation{MEMPHYS - Center for Biomembrane Physics, Department of
Physics and Chemistry, University of Southern Denmark, Campusvej 55,
5230 Odense M, Denmark}
\author{Igor M. Sokolov}
\affiliation{Institut f{\"u}r Physik, Humboldt Universit{\"a}t
zu Berlin, Newtonstra{\ss}e 15, 12489 Berlin, FRG}
\author{Ralf Metzler}
\affiliation{Physik Department, Technical University of Munich,
James Franck Strasse, 85747 Garching, Germany}


\begin{abstract}
We consider the effective surface motion of a particle that freely diffuses
in the bulk and intermittently binds to that surface. From an exact approach
we derive various regimes of the effective surface motion characterized by
physical rates for binding/unbinding and the bulk diffusivity. We obtain a
transient regime of superdiffusion and, in particular, a saturation regime
characteristic for the cylindrical geometry. This saturation, however, in a
finite system is not terminal but eventually turns over to normal surface
diffusion. The first passage behavior of particles to the cylinder surface 
is derived. Consequences for actual systems are discussed.
\end{abstract}

\maketitle

\begin{figure}
\includegraphics[width=7.2cm]{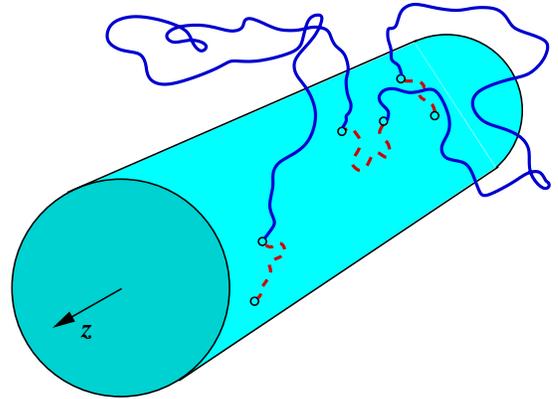}
\caption{A particle diffuses in the bulk (full lines) and intermittently
binds to a surface on which it can also diffuse (broken lines). This
produces an effective surface motion.}
\label{scheme}
\end{figure}

Bulk mediated surface diffusion (BMSD) defines the effective surface motion
of a particle on a reactive surface that intermittently unbinds and diffuses
in the adjacent bulk before rebinding (Fig.~\ref{scheme}).
BMSD was revealed by NMR in porous glasses \cite{stapf} and has relevance
to numerous technological applications \cite{bychuk1}. The particular case
of BMSD on a cylindrical surface is of
importance for facilitated diffusion in gene regulation \cite{bvh,michael},
the net motion of
motor proteins along cytoskeletal filaments \cite{motor}, the transient
binding of chemicals to nanotubes \cite{nanotubes}, or the exchange behavior
between cell surface and surrounding bulk of rod-shaped bacteria
(bacilli) and their linear arrangements \cite{bacillus} to name but a few
examples.

BMSD was previously investigated for a planar surface in terms of scaling
arguments \cite{bychuk,bychuk1}, master equation schemes \cite{revelli},
and simulations \cite{fatkullin}. More recently the first passage
problem between particle unbinding and rebinding for a free cylindrical
surface was considered \cite{levitz}. Here we establish an exact treatment
of BMSD for a reactive cylindrical surface deriving explicit expressions for
the surface occupation, the effective mean squared displacement (MSD) on the
surface, and the returning time distribution from the bulk. Different regimes
ranging from transient superdiffusion to terminal normal diffusion emerge
naturally from the physical timescales entering our description.
We derive previously unknown regimes characteristic of the cylindrical
geometry, most remarkably the existence of extremely long jumps as well as
a saturation regime of the
surface MSD that will be crucial to fully appreciate effective surface motion
mediated by bulk diffusion and its experimental investigation.

We consider a cylinder with radius $a$ centered along the symmetry axis of
an outer cylinder of radius $b$. As we are interested in the effective
displacement along the cylinder in $z$ direction (Fig.~\ref{scheme}) we
assume rotational symmetry around the $z$ axis. The concentration of
particles (density function, DF) in the bulk is denoted by $C(r,z,t)$ ($[C]=
\mathrm{cm}^{-3}$) while $n(z,t)$ ($[n]=\mathrm{cm}^{-1}$) refers to the
particle DF
per length on the inner cylinder surface. The particle numbers are
$N_b(t)=2\pi\int_a^brdr\int_{-\infty}^{\infty}C(r,z,t)dz$ in the bulk and
$N_s(t)=\int_{-\infty}^{\infty}n(z,t)dz$ on the cylinder surface. With
surface and bulk diffusivities $D_s$ and $D_b$
the two densities evolve according to the diffusion equations
(a derivation from a discrete model will be presented elsewhere
\cite{bmsdlong})
\begin{equation}
\label{desurf}
\frac{\partial}{\partial t}n(z,t)=D_s\frac{\partial^2}{\partial z^2}
n(z,t)+2\pi aD_b\left.\frac{\partial C}{\partial r}\right|_{r=a}
\end{equation}
in axial direction on the cylinder and
\begin{equation}
\label{debulk}
\frac{\partial}{\partial t}C(r,z,t)=D_b\left(\frac{1}{r}\frac{\partial}{
\partial r}r\frac{\partial}{\partial r}+\frac{\partial^2}{\partial z^2}
\right)C(r,z,t)
\end{equation}
in the bulk using cylindrical coordinates. Note the coupling term between
surface and bulk DFs through the flux term in Eq.~(\ref{desurf}). This term
is positive when the particle concentration is higher above the cylinder
surface and negative otherwise. We equip the diffusion equations with
the following boundary conditions:
\begin{equation}
\lim_{r\to a}C(r,z,t)=\mu n(z,t),\quad\mathrm{and}\quad\left.\frac{\partial
C}{\partial r}\right|_{r=b}=0.
\end{equation}
Thus right above the cylinder surface the bulk concentration is defined by the
surface density where the coupling constant $\mu=1/(2\pi ak_b\tau_{\mathrm{
off}})$ involves the binding rate $k_b$ and the mean unbinding time
$\tau_{\mathrm{off}}$. The second relation defines a reflecting condition at
$r=b$. The initial condition corresponds to a sharp concentration at the
surface:
\begin{equation}
\lim_{t\to0}n(z,t)=N_0\delta(z)\quad\mathrm{and}\quad\lim_{t\to0}C(r,z,t)=0.
\end{equation}
Averaging over the diffusion equations we readily see that $N_s(t)+N_b(t)=N_0$.

Fourier-Laplace transforming the DFs according to
\begin{equation}
C(r,k,u)=\int_{-\infty}^{\infty}dze^{ikz}\int_0^{\infty}dte^{-ut}C(r,z,t)
\end{equation}
and analogously for $n(k,u)$ (we use the explicit dependence on the Fourier
and Laplace variables to denote the transforms), the surface DF yields
in the form
\begin{equation}
\label{prop}
n(k,u)=\frac{N_0}{u+k^2D_s+\kappa q\Delta_1/\Delta}
\end{equation}
where $q=\sqrt{k^2+u/D_b}$ and $\kappa\equiv2\pi a\mu D_b=D_b/(k_b
\tau_{\mathrm{
off}})$. We thus find in the denominator of the usual diffusion propagator a
correction proportional to $\kappa$ stemming from the bulk exchange. For the
bulk DF we obtain
\begin{equation}
C(r,k,u)=\mu N_0\frac{K_1(qb)I_0(qr)+I_1(qb)K_0(qr)}{\Delta\left[u+D_sk^2+
\kappa q\Delta_1/\Delta\right]}
\end{equation}
where the $\Delta$-factors are defined as follows:
\begin{eqnarray}
\nonumber
\Delta_1&=&K_1(qa)I_1(qb)-I_1(qa)K_1(qb),\\
\Delta&=&I_0(qa)K_1(qb)+I_1(qb)K_0(qa)
\label{bessels}
\end{eqnarray}
in terms of the modified Bessel functions.

We note that the coupling strength $\kappa$ is a measure
for the efficiency of the bulk-surface exchange. At $\kappa=0$ the number of
surface particles $N_s(t)$ remains constant. We define the coupling time
scale
\begin{equation}
t_{\kappa}=D_b/\kappa^2=[k_b\tau_{\mathrm{off}}]^2/D_b
\end{equation}
that tends to infinity for vanishing coupling and to zero for strong
coupling ($\kappa\to\infty$). From the diffusion behavior we extract two
additional timescales, namely
\begin{equation}
t_a=a^2/D_b\quad\mathrm{and}\quad t_b=b^2/D_b.
\end{equation}
For times shorter than $t_a$ the particle does not yet sense the curvature of
the cylinder surface while for times longer than $t_b$ it feels the confinement
by the outer cylinder. In what follows we are mainly interested in the regime
of strong coupling, $t_{\kappa}\ll t_a\ll t_b$ but will also report an almost
ballistic behavior under weaker coupling. The remaining cases will be
discussed elsewhere \cite{bmsdlong}.

The number of adsorbed particles follows from
$n(k,u)$ by taking $k\to0$. For short times $t\ll t_{\kappa}$ we find
that $N_s(t)\approx N_0$ remains constant as it should by definition of
the coupling time. At longer times $t_{\kappa}\ll t\ll t_a$ the behavior
changes to $N_s(t)\approx N_0\sqrt{t_{\kappa}/[\pi t]}\simeq t^{-1/2}$.
For even longer times $t_a\ll t\ll t_b$ we have
a faster decay $N_s(t)\approx{\textstyle \frac{1}{2}}N_0\sqrt{t_{\kappa}t_a}
/t$ inversely proportional
to $t$. Finally at very long times $t\gg t_b$ the dynamics equilibrates
with respect to the radial diffusion and $N_s(t)\approx2N_0
\sqrt{t_{\kappa}t_a}/t_b$.

\emph{Surface diffusion.} The MSD on the cylinder surface can be obtained
from the characteristic function through
\begin{equation}
\label{surfprop}
\langle z^2(u)\rangle=-N_0^{-1}\left.\frac{\partial^2n(k,u)}{\partial k^2}
\right|_{k=0}
\end{equation}
where we divide by $N_0$ to obtain an effective one particle displacement.
This quantity is not corrected for particles leaving the surface in contrast
to the normalized displacement $\langle z^2(u)\rangle_{n}=N_0\langle z^2(u)
\rangle/N_s(t)$. For the surface MSD we find at short
times ($t\ll t_{\kappa}$)
\begin{equation}
\label{msdshort}
\langle z^2(t)\rangle\sim2D_st\left[1+\frac{2}{3\sqrt{\pi}}\frac{D_b}{D_s}
\left(\frac{t}{t_{\kappa}}\right)^{1/2}\right]\sim\langle z^2(t)\rangle_n.
\end{equation}
In this regime, that is, the diffusion to leading order is confined to the
surface and exchange with the bulk leads to a higher order correction, when
$D_s\ll D_b$, that invokes $\sim t^{3/2}$ superdiffusion (see below).
Conversely, for longer times $t_{\kappa}\ll t\ll t_a$, $N_s(t)$ decays
perceptibly and
\begin{subequations}
\begin{eqnarray}
\langle z^2(t)\rangle&\sim&2D_st_{\kappa}+\frac{2}{\sqrt{\pi}}D_b\sqrt{tt_{
\kappa}},\\
\langle z^2(t)\rangle_n&\sim&2\sqrt{\pi}D_s\sqrt{tt_{\kappa}}+2D_bt
\end{eqnarray}
\end{subequations}
At even longer times $t_a\ll t\ll t_b$  we see the influence of the cylindric
geometry in the logarithmic dependencies
\begin{subequations}
\begin{eqnarray}
\langle z^2(t)\rangle&\sim&\frac{D_st_at_{\kappa}}{t}\log\left(\frac{4t}{Ct_a}
\right)+D_b\sqrt{t_at_{\kappa}},\\
\langle z^2(t)\rangle_n&\sim&2D_s\sqrt{t_at_{\kappa}}\log\left(\frac{4t}{Ct_a}
\right)+2D_bt
\end{eqnarray}
\end{subequations}
where $\log C=\gamma=0.5772$ is Euler's constant. In this regime $\langle z^2
(t)\rangle$ thus exhibits a saturation unique to the cylinder case.
Finally, at very long times $t\gg t_b$ due to radial equilibration a linear
diffusive behavior yields
\begin{subequations}
\begin{eqnarray}
&&\langle z^2(t)\rangle\sim\frac{8t_at_{\kappa}}{t_b^2}D_st+4D_b\frac{\sqrt{
t_at_{\kappa}}}{t_b}t\\
&&\langle z^2(t)\rangle_n\sim4D_s\frac{\sqrt{t_at_{\kappa}}}{t_b}t+2D_bt,
\end{eqnarray}
\end{subequations}
i.e., the combination of bulk and surface diffusion gives rise to an
effective diffusivity involving all time scales.

\begin{figure}
\includegraphics[width=8.8cm]{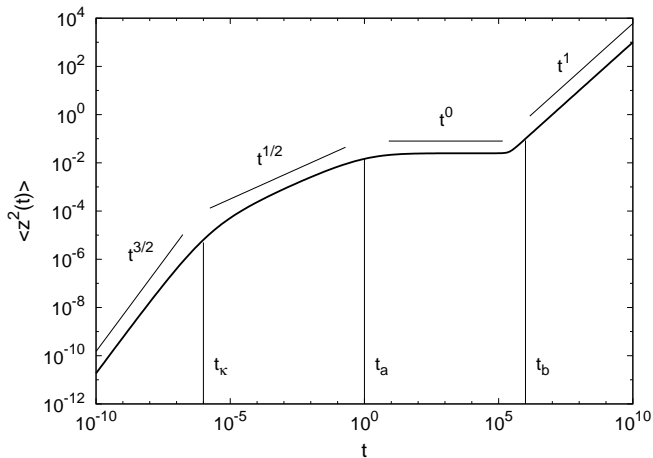}
\caption{Mean squared displacement showing the various effective surface
diffusion regimes. Note the transient plateau.}
\label{msd}
\end{figure}

The behavior of the surface MSD $\langle z^2(t)\rangle$ in absence of surface
diffusion ($D_s=0$) is shown in Fig.~\ref{msd} obtained from numerical Laplace
inversion of Eqs.~(\ref{surfprop}), (\ref{prop}), and (\ref{bessels}). Note
that the time scales (in dimensionless units) $t_{\kappa}=10^{-6}$, $t_a=1$,
and $t_b=10^6$ were chosen to be well separated, to visualize the four
scaling regimes. The other parameters were $a=5$, $D_b=a^2/t_a=25$, $\kappa=a/
\sqrt{t_at_{\kappa}}=5\times10^3$, and $b=a\sqrt{t_b/t_a}=5\times10^3$. 

\emph{Surface propagator.} Our formalism also produces the exact distributions
$C(r,t)$ and $n(z,t)$. For times longer than $t_b$ we already found that the
diffusion behavior is normal albeit with rescaled diffusivity and the surface
propagator turns into a Gaussian. We recover a Gaussian for $n(z,t)$ in absence
of bulk coupling, $\kappa=0$. For times shorter than $t_b$ and $\kappa>0$
we observe deviations from the Gaussian form. In most relevant cases $D_s\ll
D_b$ and we therefore consider the case $D_s=0$ in what follows.

At shorter times $t\ll t_a<t_b$ we have $ut_a\gg 1$ and $ut_b\gg 1$ and
$qa=a\sqrt{k^2+u/D_b}\ge a\sqrt{u/D_b}=\sqrt{ut_a}\gg1$ such that $qa\gg1$
and $qb\gg1$. Therefore $\Delta_1\approx\Delta$ and the propagator
reduces to
\begin{equation}
n(k,u)=\frac{N_0}{u+\kappa\sqrt{k^2+u/D_b}}.
\end{equation}
In the central part of the DF where $k^2>u/D_b$ (i.e., $z<\sqrt{D_bt}$), we
obtain the normalized Cauchy distribution
\begin{equation}
n(k,u)=\frac{N_0}{u+\kappa|k|}\,\,\Leftrightarrow\,\,n(z,t)=\frac{N_0\kappa t}
{\pi\left(z^2+\kappa^2t^2\right)}.
\end{equation}
This interesting result is analogous to the findings from Ref.~\cite{bychuk1}
for a flat surface found from involved scaling arguments. In contrast, here
we derive the Cauchy law from an exact approach allowing us to study the
transition to other regimes explicitly. To do so we introduce the range of
validity $\ell_C(t)=\sqrt{D_bt}$ of the Cauchy region. While at distances
$z>\ell_C(t)$ we observe a Gaussian cutoff,
for $z<\ell_C(t)$ the Cauchy approximation is valid.
From this Cauchy part we obtain the superdiffusive contribution
\begin{equation}
\int_{-\ell_C(t)}^{\ell_C(t)}\frac{z^2\kappa t\,dz}{\pi\left(z^2+\kappa^2t^2
\right)}\sim2\kappa\sqrt{D_b}t^{3/2}
\end{equation}
to the MSD, consistent with the exact result (\ref{msdshort}) [with $D_s=0$].
Calculation of the MSD from the full Eq.~(\ref{prop}) however
requires the $k\to0$ limit and thus the extreme wings of the distribution are
explored. As the system
evolves in time the central Cauchy part spreads.
As already in the regime $t_{\kappa}<t<t_a$ we have $D_bt<\kappa^2t^2$ the
asymptotic behavior $\simeq z^{-2}$ can no longer be observed.

At intermediate times $t_{\kappa}\ll t\ll t_a$ we find $n(k,u)=N_0\sqrt{t_{
\kappa}/(u+D_bk^2)}$ corresponding to the Gaussian
\begin{equation}
\label{propint}
n(z,t)=N_s(t)\sqrt{\frac{1}{4\pi D_bt}}\exp\left(-\frac{z^2}{4D_bt}\right)
\end{equation}
with $N_s(t)=\int_{-\infty}^{\infty}n(z,t)dz=N_0\sqrt{t_{\kappa}/(\pi t)}$. For
longer times $t_a\ll t\ll t_b$ the propagator acquires the shape $n(k,u)=-aN_0
\log\left(C^2[a^2k^2+ut_a]/4\right)/[2\kappa]$ which yields Eq.~(\ref{propint})
with $N_s(t)=N_0\sqrt{t_at_{\kappa}}/[2t]$. Finally, for very long times $t\gg
t_b$ we again find Eq.~(\ref{propint}) with the saturation value $N_s(t)=2N_0
\sqrt{t_at_{\kappa}}/t_b$. Indeed one can show by the shift theorem of the
Laplace transformation that any function of the argument $u+D_bk^2$ will lead
to the Gaussian shape (\ref{propint}), with appropriate normalization.

\emph{First passage.} The DF of times $\wp(t)$ a particle spends in the
bulk after detachment from the cylinder can be calculated explicitly (details
of the calculation will be presented elsewhere \cite{bmsdlong}). To this end
we initially place
the test particle at radius $a<r_0<b$ and calculate when it is first adsorbed
at $r=a$. With $t_0=r_0^2/D_b$ this first passage problem defines $\wp(t)$ by
\begin{equation}
\wp(t)=2\pi a\int_{-\infty}^{\infty}D_b\left.\frac{\partial C(r,z,t)}{
\partial r}\right|_{r=a}dz
\end{equation}
in terms of the radial flux into $r=a$. The Laplace transform of $\wp(t)$
then becomes
\begin{equation}
\wp(u)=\frac{I_1\left(\sqrt{ut_b}\right)K_0\left(\sqrt{ut_0}\right)+K_1
\left(\sqrt{ut_b}\right)I_0\left(\sqrt{ut_0}\right)}{I_1\left(\sqrt{ut_b}
\right)K_0\left(\sqrt{ut_a}\right)+K_1\left(\sqrt{ut_b}\right)I_0\left(
\sqrt{ut_a}\right)}.
\end{equation}
For $t_b\to\infty$, $\wp(u)\sim K_0\left(\sqrt{
ut_0}\right)/K_0\left(\sqrt{ut_a}\right)$ while for $r_0=a$ we recover the
sharp form $\wp(t)=\delta(t)$ as it should.

At shorter times $t\ll t_a<t_0$ we obtain the expansion
\begin{equation}
\wp(t)=\sqrt{\frac{a}{r_0}}\frac{r_0-a}{\sqrt{4\pi D_bt^3}}
e^{-\frac{\left(r_0-a\right)^2}{4D_bt}}\left(1+\frac{D_bt}{
4ar_0}+\ldots\right).
\end{equation}
This is to leading order the first passage DF for a one-dimensional random
walk reweighted by the ratio $\sqrt{a/r_0}$. Keeping the distance $r_0-a$
fixed but letting both $r_0$ and $a$ tend to infinity we recover the result
for a flat surface for which the 1D first passage remains valid at all times.

At longer times $t_a\ll t\ll t_b$ the logarithmic form $\wp(u)\sim1-2\log(r_0
/a)/\log(1/[ut_a])$ yields.
From Tauberian theorems \cite{havlin} we infer the first passage behavior
into a cylinder of radius $a$,
\begin{equation}
\wp(t)\sim\frac{2\log(r_0/a)}{t\log^2(t/t_a)},
\end{equation}
see also Refs.~\cite{levitz,redner}.
We note that a distribution of return times to the cylinder of this form
implies by a diffusive coupling $z^2\simeq t$ that a single bulk excursion
leads to the DF $\lambda(z)\simeq1/(z\log^2z)$ of effective dislocations
$z$ along the cylinder. Finally in the long time limit $t\gg t_b$ the outer
cylinder comes into play and $\wp(t)$ attains an exponential cutoff, leading
to a mean first passage time
\begin{equation}
\langle t\rangle=b^2\log\left(r_0/a\right)/[2D_b].
\end{equation}

\emph{Extremely long jumps.} An interesting behavior occurs when we relax
the strong coupling condition $t_{\kappa}\ll t_a$. We said that when $t
\gg t_a$ the diffusing particles feels a change in geometry from planar to
cylindrical, and the time scale $t_c\equiv\sqrt{t_a t_\kappa}=a/\kappa$
occurs in our expressions.
If we have $t_a\ll t\ll t_c\ll t_b$ then $N_s(t)\approx N_0$ and
\begin{equation}
\left<z^2(t)\right>=2D_st+\frac{2D_bt^2}{t_c\log^2\left[4t/(C^2 t_a)\right]},
\end{equation}
a ballistic behavior with logarithmic correction: For $D_s \ll D_b$ the
superdiffusion is even stronger than for $t\ll t_a$. Similarly,
the propagator reduces to
\begin{equation}
n(k,u)=N_0/[u+\kappa q K_1(a q)/K_0(a q)].
\end{equation}
In the range $a \ll z \ll \sqrt{D_b t}$ (i.e., $\sqrt{u t_a} \ll a k \ll 1$),
$\kappa q K_1(a q)/K_0(a q)\sim (\kappa/a) \log[2/(C a k)]$, an extremely
weak $k$ dependence: this pre-cutoff tail of $n(k,u)$ is
heavier than any normalizable power law, due to the
heavy tailed distribution of bulk mediated dislocations.

\emph{Discussion.} We established an exact approach to BMSD on a reactive
cylindrical surface revealing four distinct surface diffusion regimes. In
particular our formalism provides a stringent derivation of the transient
superdiffusion discussed earlier \emph{and\/} explicitly quantifies the
transition to other regimes. Notably we revealed a saturation regime for
the surface MSD that becomes relevant at times above which the diffusing
particle feels the curvature of the cylinder surface ($t_a$). This behavior,
caused by the cylindrical geometry,
stems from an interesting balance between a net flux of particles into the
bulk and the fact that particles with a longer return time also lead to an
increased effective surface relocation. In absence of an outer cylinder the
saturation is terminal, while in its presence the surface MSD returns to a
linear growth in time. This observation will be important in future models
of BMSD around cylinders and particularly for the interpretation of
experimental data obtained for BMSD systems. We note that in the proper
limit $a\to\infty$ the previous results for a planar surface are recovered.
Relaxing the strong coupling condition we demonstrated the existence of an
almost ballistic surface diffusion behavior, a case that might be relevant
for transport along thin cylinders such as DNA.

In Ref.~\cite{levitz} it was shown that the scaling behavior in the
regimes below and above $t_a$ can be probed experimentally by NMR methods
measuring the BMSD of water molecules along imogolite nanorods over three
orders of magnitude in frequency space. For larger molecules such as a
protein of approximate diameter 5 nm we observe a diffusivity of $10^{-6}
\mathrm{cm}^2/\mathrm{sec}$ such that for instance the saturation plateau
around a bacillus cell (radius 1/2 $\mu$m) sets in at around $t_a=2.5$ msec
which might give rise to interesting consequences for the material exchange
around such cells. In general, the relevance of the individual regimes will
crucially depend on the scales of the surface radius and the diffusing
particle (and therefore its diffusivity). It was discussed previously that
even the superdiffusive short-term behavior may become relevant
\cite{bychuk,bychuk1,fatkullin}. In general, in a given system the separation
between the various scaling regimes may not be sharp. Moreover typically a
single experimental technique will not be able to probe all regimes. It is
therefore vital to have available a solution for the entire BMSD problem.

\end{document}